\documentclass[preprint,12pt]{elsarticle}

\usepackage{epsfig}

\usepackage{amssymb}

\usepackage[ps2pdf,%
a4paper=true,%
breaklinks=true,%
colorlinks=true,%
pdfauthor={Rene A. Ong},%
pdftitle={Highlights from VERITAS on VHE Gamma-ray Sources in our Galaxy}%
]{hyperref}

\journal{Advances in Space Research}

\begin{document}

\begin{frontmatter}



\title{Highlights from VERITAS on VHE Gamma-ray Sources in our Galaxy}


\author{Rene A. Ong, for the VERITAS Collaboration}
\address{Department of Physics and Astronomy, University of California, Los Angeles, 
         Los Angeles, CA, 90095-1547, USA}




\begin{abstract}

VERITAS is a major ground-based detector of very high energy (VHE, E $>100$ GeV) gamma rays and cosmic rays. VERITAS consists of an array of four 12m-diameter atmospheric Cherenkov telescopes that has been fully operational since September 2007. VERITAS has detected many astrophysical sources of VHE gamma rays, including at least 17 VHE sources that are likely 
Galactic in origin.
This paper describes some of the Galactic source highlights from VERITAS with an emphasis on those aspects that relate to 
the origin of cosmic rays.
Specifically, topics include the VERITAS discovery of VHE emission from 
the Tycho and CTA 1 supernova remnants,
the identification of HESS J0632+057 as a new VHE binary, a substantially improved view of the gamma-ray 
emission in the Cygnus OB1 region, and the recent remarkable discovery of VHE emission from the Crab pulsar.
In 2009, VERITAS was upgraded by relocation of one of the telescopes, leading to a significant improvement in 
sensitivity. A program to further improve the performance of VERITAS, involving the upgrade of the telescope 
trigger systems and the telescope cameras, was completed in summer 2012. The upgrade will lead to an improved 
sensitivity and a
lower energy threshold 
for VERITAS, allowing it to perform deeper observations 
of known Galactic and extragalactic sources and to detect fainter and more distant sources. 

\end{abstract}


\end{frontmatter}

\parindent=0.5 cm

\section{Introduction}

This year, we celebrate the 100th anniversary of the discovery of cosmic rays.
These particles, mostly charged, span an enormous energy range from below 1 GeV
to above $10^{20}$\,eV.
They pervade our Galaxy with an energy density 
($\sim 1$ eV/cm$^3$) comparable to that in the Galactic magnetic field or
the cosmic microwave background.  Yet, after many years of research,
we still do not fully understand the origin of the cosmic rays above
100\,GeV, the so-called very high-energy (VHE) gamma-ray regime.
Neutral messengers, i.e. gamma rays and neutrinos, can provide critical
information about cosmic ray origins because they directly trace the
energetic processes taking place at acceleration sites in our Galaxy
and beyond.

VHE gamma ray telescopes, such as VERITAS, have already detected a rich
variety of sources that can shed light on the origin of cosmic rays.
First and foremost are numerous detections of supernova remnants (SNRs),
objects that are generally believed to produce cosmic rays up to the
knee in the energy spectrum (i.e. E $\sim 10^{15}\,$eV) through the
mechanism of diffuse shock acceleration.
Other detected VHE sources in our Galaxy include:  
1) binary systems, in which the acceleration may involve accretion-powered jets
or colliding winds,
2) pulsars and pulsar nebulae (PWNe), in which the spin down of the neutron star
powers a relativistic wind of particles,
and
3) star forming regions, in which the gamma rays trace the interactions
taking place in molecular clouds or near OB associations.
It is important to note that many of the VHE gamma-ray sources in the Galactic
plane have yet to be firmly associated with any known astronomical object.
Understanding the nature of these unidentified objects is thus an important
goal for the field, especially in the context of the complementary information
provided at $\sim 1\,$GeV by the Fermi space telescope.

VHE gamma rays have important advantages in cosmic ray origin studies because
they are not deflected by interstellar magnetic fields and they trace the energetic
particles accelerated by shocks (and other mechanisms) at the sources.
In addition, unlike the situation at $\sim 1\,$GeV, there is not a dominant
Galactic diffuse background at very high energies.
However, a difficulty associated with gamma rays is that
they can be produced from the interactions of both high-energy protons and electrons.
Protons (and nuclei) produce gamma rays through their interaction with target
material (e.g. a molecular cloud) and subsequent $\pi^\circ$ decay.
Electrons (and positrons) produce gamma rays through the processes of
inverse-Compton scattering and bremsstrahlung.
Since the gamma rays come from secondary interactions for both types of primary
particles, they trace the combination of the beam (primary particle) density
and the target density.
For any particular VHE source, it is likely that {\em both} protons and
electrons are being accelerated and the need to disentangle the 
two components often requires additional information derived
from multi-wavelength
observations.

In this paper, we discuss recent results from the VERITAS observatory that
are of particular interest to the question of high-energy cosmic rays.
We introduce VERITAS and then discuss its observations of Galactic sources, including
SNRs such as Tycho and CTA 1, the newly-identified 
binary system HESS J0632+057, various sources in the Cygnus region of the
Galaxy, and the Crab pulsar.
We conclude with a description of the just-completed VERITAS upgrade
and provide expectations for the future science program of VERITAS in this area.

\begin{figure}
\begin{center}
\includegraphics*[width=13cm]{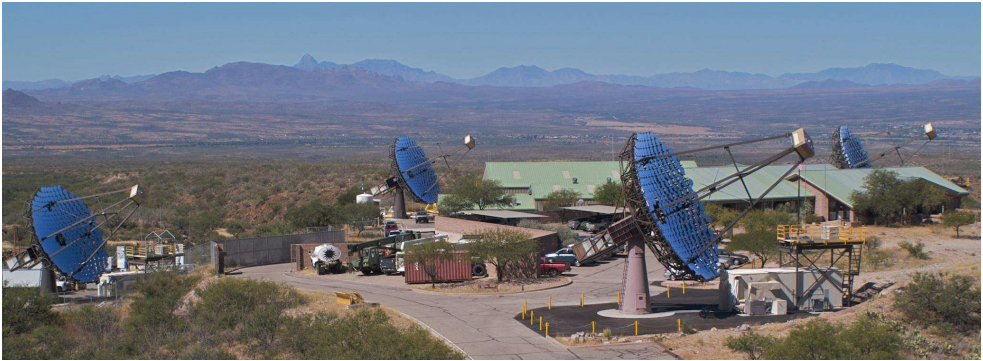}
\end{center}
\caption{The VERITAS array of four atmospheric Cherenkov telescopes in southern Arizona, USA.}
\end{figure}

\section{VERITAS}

VERITAS (the Very Energetic Radiation Imaging Telescope Array System) is a state-of-the-art
ground-based VHE gamma-ray observatory.
Located at the F.L. Whipple Observatory in southern Arizona, USA, VERITAS
uses the imaging atmospheric Cherenkov technique to detect gamma rays at energies from
$\sim 50\,$GeV to more than 20\,TeV. 
The observatory consists of an array of four large telescopes (see Figure~1), each consisting of
a 12\,m-diameter optical reflector and an associated camera spanning a field-of-view
of 3.5$^\circ$.
Each telescope produces an image of the Cherenkov light generated in the atmosphere.
A stereo reconstruction technique makes use of the orientation, intensity and 
shape of each telescope image to estimate the direction, energy, and type 
 of the primary particle (i.e. gamma ray or hadron).

Standard observations with the full four-telescope VERITAS array started in
September 2007 and the instrument has been upgraded successively since then.
In 2009, one of the telescopes was moved to give a more symmetrical array geometry.
The new geometry, combined with a reduced optical point spread function,
led to a significant improvement in the point-source sensitivity of VERITAS.
The VERITAS upgrade, carried out between 2009 and 2012, led to a lowering
of the energy threshold and a further improvement in the sensitivity. More details
of the upgrade are found in Section~8.

\begin{figure}
\begin{center}
\includegraphics*[width=12.5cm]{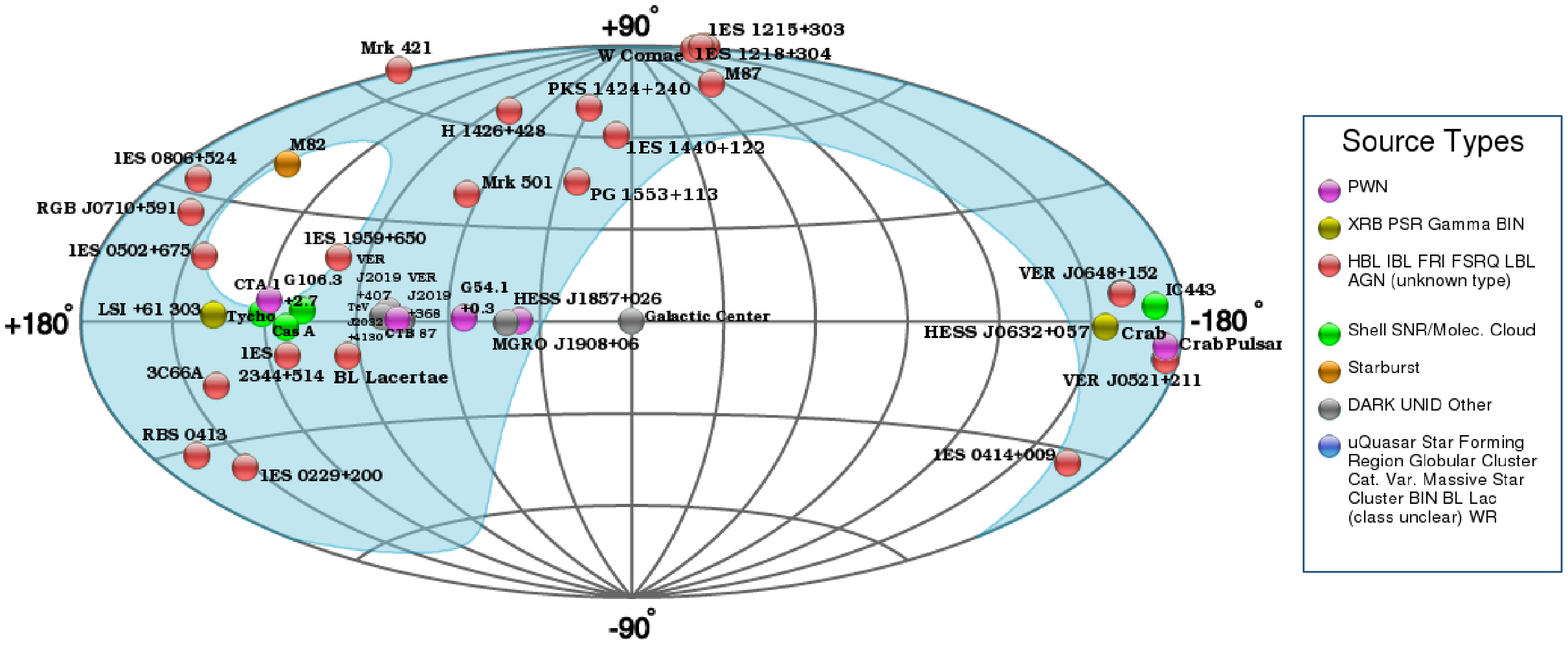}
\end{center}
\caption{The VERITAS VHE sky map. This figure shows the sources detected
by VERITAS, as of July 2012.  The source classification is shown by the legend
on the right. Plot was made using TeVCAT at http://tevcat.uchicago.edu/.}
\end{figure}

Most of the data used in the results presented here were taken between 
September 2009 and June 2012. For this period of time, VERITAS had an angular
resolution of $0.1^\circ$ at 1 TeV, an energy resolution of 15\% at 1 TeV, and 
an energy threshold of $\sim$100 GeV at the trigger level (assuming a
E$^{-2.4}$ differential spectrum).
A gamma-ray point source with a flux level of 1\% of the Crab nebula can
be detected at the 5 standard deviation (5$\sigma$) level in less than 30 hours
of observing time \cite{VERITAS-web}.

The VERITAS collaboration consists of $\sim$100 scientists from 23 institutions
in five countries (U.S., Canada, U.K., Ireland, and Germany). An additional
35 Associate members participate in the VERITAS science program and primarily use 
other instruments (e.g. IceCube, Fermi, and Swift) or come from the theoretical
and broader multiwavelength communities.
Joint science activities between VERITAS and IceCube and between VERITAS and
Fermi-LAT are carried out under the guidance of memorandums of understanding
(MOU's).  A new program (starting in 2013) allows researchers from
around the world to propose for VERITAS observation time in support of
Fermi guest investigator activities \cite{Fermi-GI}.

\section{VERITAS Sky Map}

The VERITAS sky map -- showing the sources detected by VERITAS as of
July 2012 -- is shown in Figure~2.
At least 40 VHE sources have been detected by VERITAS; each source
was detected with a post-trials statistical 
significance of greater than 5.0 standard deviations.
As shown in the figure legend, the VERITAS sources cover seven source
classes and at least 17 of them are likely to be Galactic in origin.

\begin{figure}
\begin{center}
\includegraphics*[width=9.2cm]{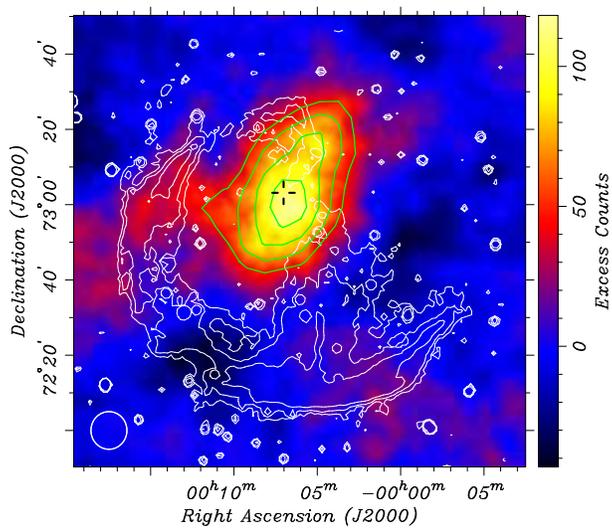}
\end{center}
\caption{VERITAS image of the supernova remnant CTA 1 \cite{Aliu2012a}.
The color scale indicates the VHE gamma-ray excess counts.
The VERITAS significance contours at 3, 4, 5, and 6$\sigma$ are shown in
green. The cross marks the position of the pulsar detected by
Fermi-LAT \cite{Abdo2008}.
The SNR shell is indicated by the radio contours (1420 MHz), shown in white.
The white circle at the bottom left indicates the size of the point-spread
function.}
\end{figure}

\section{Supernova Remnants and Pulsar Wind Nebulae: Tycho and CTA 1}

Tycho's SNR is an important historical remnant from a type 1a supernova 
first observed in 1572.
It has long been considered 
(e.g. from the X-ray morphology \cite{Warren2005})
a good candidate for demonstrating hadronic acceleration.
Based on a large data set of observations taken between 2008 and 2010, VERITAS
discovered VHE gamma-ray emission from Tycho \cite{Acciari2011a}.
The VERITAS source name for this object is VER J0025+641.
The detected flux was relatively low (0.9\% of the Crab nebula flux), making Tycho one of the
weakest VHE gamma-ray sources ever detected.
The source spectrum was relatively hard, with a differential source spectral 
index of $\Gamma = 1.95\pm 0.51\pm 0.30$.
VERITAS fit the differential spectrum from 1 TeV to 10 TeV using three bins in energy.

As discussed in \cite{Acciari2011a}, the broad-band spectral energy distribution (SED) for
Tycho can be adequately fit by both leptonic and hadronic acceleration models.
However, a recent result from data taken by the Large Area Telescope (LAT) instrument
on board the Fermi gamma-ray space telescope favors the hadronic interpretation.
The Fermi-LAT result \cite{Giordano2012} reports the detection of Tycho from
gamma-ray energies of 100\,GeV all the way down to 400\,MeV.
The flat spectrum of the gamma rays extending from the sub-GeV region (Fermi) to 
10 TeV (VERITAS) means that leptonic models, incorporating inverse-Compton and
bremsstrahlung emission, have difficulty in explaining the majority of the observed
gamma-ray flux.
Conversely, a relatively simple hadronic model, in which a population
of accelerated protons produces high-energy and VHE gamma rays via $\pi^\circ$ decay,
is able to fit the measured SED well.

CTA 1 is a composite/shell-type supernova remnant with an 
X-ray filled radio shell of $\sim 1.8^\circ$ diameter.
With an age of approximately 13 kyr and a distance of $1.4\pm 0.3\,$kpc,
CTA is a well-known radio SNR, although with no known pulsar prior to 2008.
Fermi-LAT observations led to the discovery of a pulsar in 2008
\cite{Abdo2008} with a pulsar period of 316 ms, a cutoff energy
of $\sim 5\,$\,GeV and a characteristic pulsar age that is comparable to the
age of the SNR.
The Fermi-LAT result, the first gamma-ray pulsar discovered via 
a blind-search technique,
was followed up by the detection of a X-ray pulsar by Chandra \cite{Caraveo2010}.
Although the properties of CTA 1 argue for the presence of a pulsar wind nebula (PWN),
prior to 2011 there was no specific evidence to support this picture.
 
The definitive discovery of a pulsar wind nebula associated with CTA 1 was made
by VERITAS in 2011 \cite{McArthur2011}.
The VERITAS discovery of VHE gamma-ray emission from CTA 1 was made using
approximately 40 hours of data taken in 2010 and 2011, which yielded 
a $6.5\sigma$
excess over background and an integral gamma-ray flux (above 1 TeV) of
$F_\gamma = 4.0 \times 10^{−12}\,$erg\,/cm$^{2}$/s,
corresponding to $\sim 0.2$\% of the spin-down power \cite{Aliu2012a}.
The VERITAS source name for this object is VER J0006+729.
As shown in Figure~3, the VHE gamma-ray emission is clearly extended with a spatial
size of approximately $0.25^\circ-0.30^\circ$.
The compact extension of the gamma-ray emission detected by VERITAS, 
plus the fact that the emission is centered on the location of the Fermi-LAT 
pulsar, argues for a PWN origin.
Indeed, a fit to the broadband SED of CTA 1 is consistent with
synchrotron and inverse-Compton emission from a PWN \cite{Aliu2012a}.
A comparison of CTA 1 with other X-ray and
gamma-ray pulsar wind nebulae indicates that it has a spin-down luminosity that
is typical for a pulsar of its age.

\begin{figure}
\begin{center}
\includegraphics*[width=11.0cm]{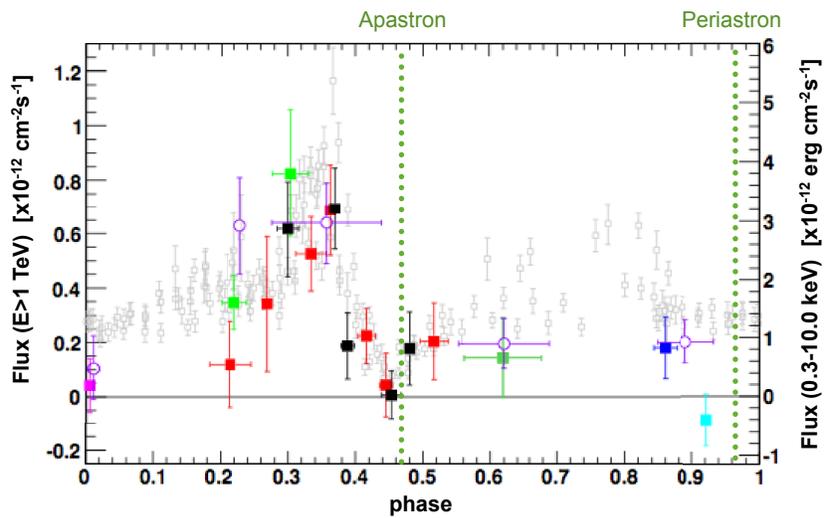}
\end{center}
\caption{Integral gamma-ray fluxes above 1 TeV of HESS J0632+057 from VERITAS (filled
markers), HESS (open markers) and Swift XRT X-ray measurements (0.3-10 keV) folded with the orbital period of 315 days
\cite{Maier2012}.
The colors indicate the different observing periods of VERITAS: Nov 2011-Feb 2012 (black),
Dec 2010-April 2011 (red), Feb 2010-March 2010 (light green), Oct 2010 (dark blue), Jan 2009 (pink),
Dec 2008 (light blue) and Dec 2006-Jan 2007 (dark green).
The HESS observations were made between 2004 and 2010.
The times of apastron and periastron are indicated by the green dotted lines.}
\end{figure}

\section{A New VHE Binary}

There are currently seven binary systems known at high energies (detected by
Fermi-LAT) and at very high energies (detected by ground-based telescopes such
as VERITAS).
These are fascinating but complicated systems in which the emission can
involve processes associated with a jet, a stellar wind, a stellar disk,
or a pulsar, or some combination of these.
Understanding the acceleration and emission mechanisms is complicated by
many unknown factors, including, e.g.,  the source geometry, the properties of 
the wind or disk, and the unknown compact object.
Until recently, the VHE binaries numbered three: PSR B1259-63, LS 5039, and
LS I +61 303 and all of these sources had also been detected at high energies.
Three additional binaries have been seen at GeV energies: Cygnus X-1, Cygnus X-3,
and 1FGL J1018.8-5856, but none of these objects has been definitively detected
at very high energies.

The source HESS J0632+057 has been a very interesting, but puzzling, source since
its discovery by HESS \cite{Aharonian2007}. The object is possibly associated with 
MWC 148, a B0pe star, in the Monoceros loop.
The fact that it was the only 
point-like unidentified source from the original HESS Galactic plane survey led to 
speculation that HESS J0632+057 
might be a binary system \cite{Hinton2009}.
In 2009, the source was demonstrated to be variable since the
VERITAS upper limit on the VHE gamma-ray flux
(of $< 1.1$\% Crab nebula above 1\,TeV \cite{Acciari2009a}) was well below
the original source detection level (of $\sim 3$\%).
Deep observations made by VERITAS in 2009 and 2010 led to a strong detection
($> 12\sigma$) of the source and confirmed its point-like nature \cite{Maier2011}.
A later detection was made by MAGIC in which the VHE emission roughly coincided
with an increased X-ray flux \cite{Aleksic2012}.

In 2011, an important breakthrough in the understanding of the source
HESS J0632+057 was 
made by Swift with the detection of a periodicity in 
X-rays \cite{Bongiorno2011}.
The detected periodicity of $\sim 321\,$days is assumed to be associated with the
orbital motion of the system and it allowed the ground-based telescopes to target
the source during periods of elevated X-ray emission.
Extensive observation campaigns were carried out by VERITAS and HESS that
resulted in a definitive detection of correlated activity in VHE gamma rays
and in X-rays.
As shown in Figure~4, the overall data set, dominated by the VERITAS
observations, shows clear flaring activity of the source at the same orbital
phase location over two different seasons \cite{Maier2012}.
These combined observations conclusively demonstrate that HESS J0632+057
is in fact a binary system, making it the first detected binary that was motivated
by VHE observations.
Although the source is not detected by Fermi-LAT, it is hoped that continued
observations at VHE energies will lead to 
a better understanding of its nature.

\begin{figure}
\begin{center}
\includegraphics*[width=13.0cm]{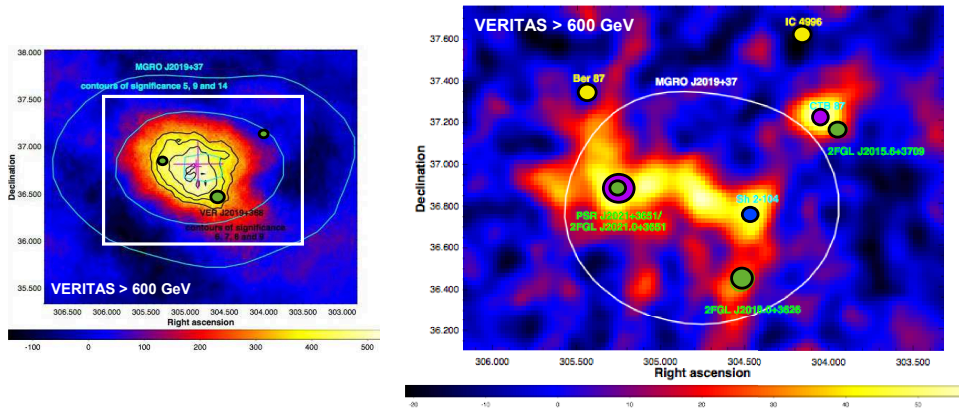}
\end{center}
\caption{VHE gamma-ray sky maps from VERITAS for the Cygnus OB1 region.
Left: excess counts at energies above 600 GeV for the overall region
observed by VERITAS (color scale).
The VERITAS significance contours of 6,7,8, and 9 standard deviations
are indicated by the black lines.
The Milagro significance contours of 5, 9, and 14 standard deviations for 
the source MGRO J2019+37 are given by the light blue lines.
Fermi-LAT sources are shown by the filled green circles.
The white rectangle shows the boundary of the blown-up region shown on the
right.
Right: blow-up of the Cygnus OB1 region showing the excess counts at
energies above 600 GeV observed by VERITAS (color scale). Fermi-LAT
sources are shown by the filled green circles. Other
known sources are shown  by the filled yellow, purple and blue circles.}
\end{figure}

\section{The Cygnus Region}

The Cygnus region of the Galactic plane contains a substantial
catalog of possible accelerators of cosmic rays:
SNRs, X-ray binaries, massive star clusters and PWNe.
This region was targeted by the Cygnus Sky Survey, a flagship project
that was carried out by VERITAS.
The base survey was carried out between 2007 and 2009 and covered a
region of $15^\circ \times 5^\circ$ in Galactic longitude and latitude,
respectively.
The survey encompassed
$\sim$150\,hours of observations at a uniform point-source sensitivity
(99\% CL) of $< 4$\% of the Crab nebula flux.  This sensitivity
is a factor of five better than previously done \cite{HEGRAsurvey}.
Follow-up observations were carried out in subsequent seasons. 
The combination of the base survey and the follow-up data 
led to the detection of a number of sources plus 
interesting hints of sources ("hotspots"), including:
1) an extended source near the shell-type SNR G78.2+2.1 
($\gamma$-Cygni), 
2) an extended source coinciding with the unidentified 
VHE gamma-ray source TeV J2032+4130, 
and 3) the Cygnus OB1 region. 
The first object represented an entirely new VHE source, named
VER J2019+407 and is discussed in more detail elsewhere
\cite{VERJ2019}.
The second object is a known gamma-ray emitter.
The VERITAS observations represent the most sensitive exposure
taken in the VHE band and are the subject of an upcoming
publication \cite{TeV2032}.
Here we discuss the third region which represents completely new
findings at these energies.

Cygnus is an active star-forming region hosting several 
OB associations, considered as tracers of young SNRs. 
Milagro discovered   
source MGRO J2019+37 in the OB1 region above 12 TeV \cite{MGROsurvey},
but the detection yielded a relatively broad (1.1$^\circ$ x 0.5$^\circ$)
feature that was not spatially resolved.
Based on this finding and
the preliminary evidence of emission from the survey,
we did deep observations ($\sim 70\,$hours)
of this region in 2010-2012, revealing 
complex VHE gamma-ray emission \cite{Aliu2011}.
The excellent angular resolution of VERITAS enables the spatial
separation of several VHE sources in the region, as shown in
Figure~5.
The region near MGRO J2019+37 likely comprises
multiple, extended sources. 
There is a separate point-like
VHE source associated with the
pulsar wind nebula CTB 87.

The possible candidates for the VHE emission near MGRO J2019+37
are a PWN associated with the energetic pulsar 
PSR J2021+3651, the Wolf Rayet star WR141, a transient INTEGRAL source 
IGR J20188+3647, and the HII region Sh104
and possibly other cosmic ray sources not yet identified.
The fact that the central, extended source region exhibits a very hard
(E$^{-1.8}$) VHE spectrum supports a picture of cosmic ray (proton)
acceleration.
By itself, CTB 87 was detected at a post-trials significance 
of 6.1$\sigma$.  
At a distance of $\sim$ 6 Kpc, CTB 87
is believed to be powered by a pulsar 
although the energetics are not yet well known. 

\begin{figure}
\begin{center}
\includegraphics*[width=12.5cm]{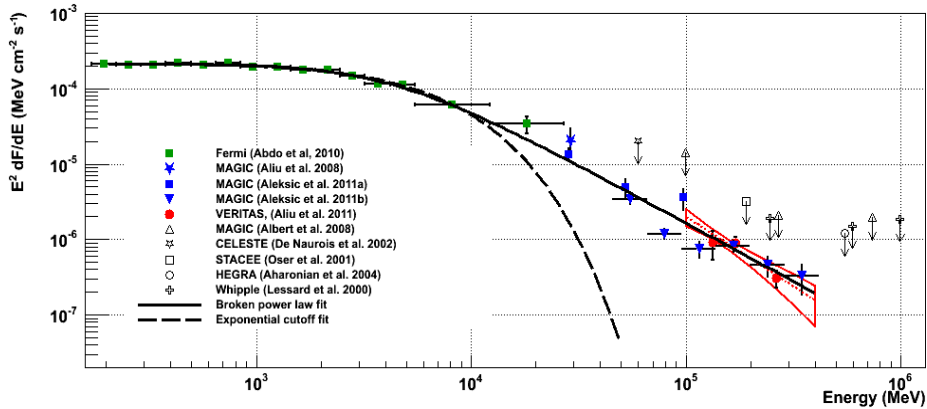}
\end{center}
\caption{High-energy and very high-energy gamma-ray spectrum
of the Crab pulsar.  The data points from the various experiments
are shown by the legend.
The red butterfly indicates the uncertainties on the published
VERITAS spectrum \cite{VERITASCrab}.}
\end{figure}

\section{Crab Pulsar}

The Crab is a famous object in astronomy
that was formed from a historical supernova observed in 1054.
The Crab nebula is the brightest, steady VHE gamma-ray source
and one that has been pivotal in the development of the field.
As one of the most energetic pulsars,
the Crab pulsar has been a continual target of observation by ground-based
gamma-ray telescopes over the last thirty years.
In the high-energy band, Fermi-LAT (and earlier EGRET) have made
exquisite measurements of the Crab pulsar spectrum showing clear
evidence for a spectral break around a few GeV.
MAGIC reported the detection of the pulsar at 25 GeV and the hint of
a signal at 60 GeV \cite{MAGICCrab}.
However, over the years, many experiments have searched for a pulsed signal above
100 GeV and have only been able to set upper limits on the emission.

Prior to 2011, the conventional view of the Crab pulsar was that the
high energy emission was described by a single component that had an
exponential cut-off in its spectrum.  The favored gamma-ray production
mechanism was curvature radiation from regions in the outer magnetosphere.
The picture dramatically changed with the announcement by VERITAS in 2011
of a pulsed signal from the Crab above 100 GeV \cite{VERITASCrab}.
The signal, derived from 107 hours of data taken between 2007 and 2009, showed
clear evidence for a two-peaked pulsed signal in the Crab phaseogram.
Also remarkable was the fact that the VERITAS Crab pulsar spectrum could be
fit by a single power-law form all the way up to 400 GeV and showed
no evidence for a cut-off, which argues for a completely new high-energy
component of emission.
The fact that gamma-ray emission was seen at these very high energies
pushed the limit on the distance of the emission region 
to beyond 10 stellar radii.
VERITAS observed that the Crab pulses are narrower in time at
very high energies compared to those observed by Fermi (and MAGIC)
at energies below 25 GeV which could imply that the acceleration region
tapers as the particle energy increases.
However, within current uncertainties, the VHE pulses occur at the same
point in phase as those at lower energies, which permits useful limits on
Lorentz invariance violation to be made \cite{Otte2012}.

Figure~6 shows the current picture of the high-energy and very high-energy
spectrum of the Crab pulsar, including a recent detection 
by MAGIC \cite{MAGICCrab2} which 
confirmed the VERITAS result above 100 GeV.
Figure~6 shows the published VERITAS spectral points.
However,
the combined VERITAS data set over the period of 2007 to 2012 now totals
130 hours and the Crab pulsed signal consists of $1514\pm 145$ excess
events, for a total statistical significance of $\sim 10.7\sigma$.
Thus, we can expect to see an improved determination of the VHE
spectrum from the Crab pulsar that will help to further constrain
theoretical models put forward to explain the unexpected emission.
It will also be important to understand if the Crab is a unique object or
if there are other pulsars with similar VHE gamma-ray behavior.

\begin{figure}
\begin{center}
\includegraphics*[width=10.2cm]{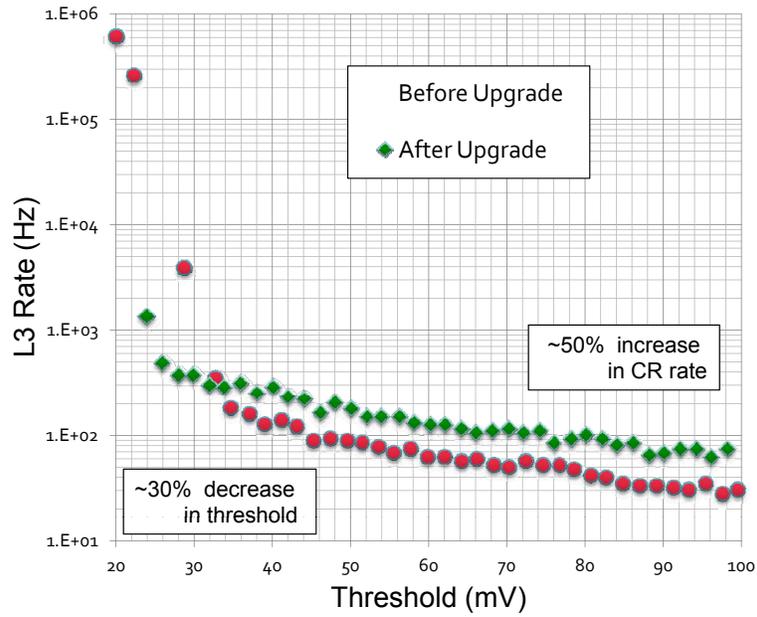}
\end{center}
\caption{VERITAS bias curves showing the Level 3 (array) trigger
rate as a function of photomultiplier tube discriminator threshold.
The red (green) points correspond to data taken before (after) the
VERITAS upgrade.  The rate at low trigger thresholds is dominated by
accidentals from night sky background. Above the break in the curve,
the rate is dominated by cosmic ray events.
One photoelectron corresponds to approximately 7.5\,mV in 
trigger threshold.}
\end{figure}

\section{VERITAS Upgrade}

VERITAS began operations with the full four-telescope array in September 2007.
Starting in 2009, we have made significant improvements to the performance of
the instrument.
The first VERITAS telescope was relocated during the summer of 2009
to improve the geometry of the array.
Combined with a superior optical point spread function resulting from
improved mirror facet alignment \cite{McCann2010}, 
the telescope relocation led to 
an improvement in both angular resolution and sensitivity --
the time to detect a point source with a flux level of 1\% of the Crab nebula
was reduced from 50 hours to $< 30$ hours.

The VERITAS upgrade (supported by NSF MRI-R2 funds)
was carried out between 2009 and 2012. 
The primary components of this program were the development and installation of
a new Level 2 (telescope) trigger system
and the replacement of all 
photomultiplier tubes (PMTs) in the VERITAS cameras by
higher quantum efficiency versions.
The trigger upgrade was carried out in 2011 and the camera upgrade was completed
during the summer of 2012, when the existing Photonis XP2970 PMTs
were replaced by Hamamatsu R10560-100 SBA PMTs.
Simulations show that the upgrade will lead to a further
improvement in sensitivity, with the time to detect a point source at
1\% of the Crab nebula flux expected to be reduced to $\sim$20 hours.
Also critical is the expected improvement in collection area at low
energies that should lead to a reduction in the energy threshold
(defined as the peak of the trigger gamma-ray events assuming a
E$^{-2.4}$ differential energy spectrum) from 100 GeV to 70 GeV.
As shown in Figure~7, data taken with the upgraded VERITAS in October 2012 
confirm the expected improvements; the trigger threshold has 
decreased by 30\% and the trigger rate has increased by 60\%.

\section{Conclusions}

VERITAS is a state-of-the-art ground-based VHE gamma-ray telescope array.
Since its start of operation in September 2007, VERITAS has produced a
wealth of new information about Galactic sources of VHE gamma rays.
A number of supernova remnants have been detected and some of the detections
provide good evidence for the acceleration of cosmic rays to very high energies.
For example, Tycho is a relatively clean system where the combined data
from VERITAS and Fermi-LAT clearly favor a hadronic acceleration picture.
Similarly, new results from VERITAS support
cosmic ray acceleration taking place in the Cygnus OB1 region of 
the Galactic plane.  VERITAS has also recently discovered VHE gamma-ray emission
from well-known supernova remnants CTA 1 and CTB 87.  The VHE emission from
these two sources most likely originates in a pulsar wind nebula.
Binary systems are relatively rare (and not fully understood) at high
energies and so the recent identification of HESS J0632+057 as both an X-ray
and VHE gamma-ray binary is an important achievement.
Finally, the unexpected discovery by VERITAS of VHE gamma-ray emission from
the Crab pulsar has led to an important re-assement of our understanding
of emission processes in pulsars and it should, in future, lead to stringent
tests of Lorentz invariance violation.
The recently completed upgrade of VERITAS is expected to lead
to more discoveries and a significantly improved understanding of high-energy
and very high-energy processes in our Galaxy.

\vspace{1.0cm}

\noindent{\bf Acknowledgements}

\vspace{0.1cm}

This research is supported by grants from the U.S. Department of Energy Office
of Science, the U.S. National Science Foundation and the Smithsonian Institution, by
NSERC in Canada, by Science Foundation Ireland (SFI 10/RFP/AST2748) and by STFC
in the U.K. 
We acknowledge the excellent work of the technical support staff at the Fred
Lawrence Whipple Observatory and at the collaborating institutions in the construction
and operation of the instrument.


\end{document}